\documentclass[twocolumn,prl,aps,showpacs,superscriptaddress,epsfig,dvips,amsmath,amssymb]{revtex4}

\usepackage{eufrak}
\usepackage[dvips]{graphicx}
\usepackage{dcolumn}
\usepackage{bm}
\usepackage{multirow}

\begin{document}
\title{A Semigroup Theory of Rate Independent Hysteresis}
\author{Xiangjun Xing}
\affiliation{Department of 
Physics, Syracuse University, Syracuse, New York~13244}
\email{xxing@physics.syr.edu}
\date{\today} 
\begin{abstract}
We explore a macroscopic, algebraic approach to rate independent hysteresis using semigroup theory.    A macroscopic description of metastable states relevant to rate independent hysteresis is introduced using field history.  The semigroup structure of the history space is identified.  Using semigroup theory and related mathematical techniques, the general relation between return point memory (RPM) and partial order is discovered.  For hysteresis system with RPM, a variational principle is identified.  The erasing properties of field histories are also characterized.   The connection between this semigroup approach and other models are also discussed.  
\end{abstract}

\pacs{75.60.Ej, 03.65.Fd, 02.90.+p}
\maketitle

\noindent
Rate independent hysteresis refers to irreversible phenomena in the adiabatic limit.  It is universally exhibited by disordered, frustrated systems at low temperature, such as random magnets, glasses, plastic materials, etc.  These systems typically have exponentially large number of metastable states.   When thermal fluctuations are negligible, such a system stay indefinitely in whatever metastable state it happens to find.  Consequently, the system properties depend on the whole history of external field back to infinite past.   

Theories of rate independent hysteresis \cite{book:Mayergoyz,book:Bertotti,book:Visintin,book:Hill} has been studied traditionally by the engineer community.  Engineers have long understood that bi-stability (or rather multi-stability) is the essential ingredient of hysteresis, and have proposed a wide range of models \footnote{For a review of different models on rate independent hysteresis, see reference \cite{book:Visintin}.} to account for these phenomena, such as the play model, the stop model, the Preisach model \cite{ref:Preisach}, the Stoner-Wohlfarth model \cite{ref:Stoner}.  These models are usually based on various simple microscopic mechanisms and provide definite schemes to calculate hysteresis loops.  Mathematicians have also came up with models involving partial differential equations \cite{book:Visintin}.   


The physics of zero temperature disordered system has also attracted considerable interests in physics community in recent years.  One primary interests seems to be the zero temperature critical phenomena exhibited by hysteresis loops of random field Ising model (RFIM) \cite{RFIM-review,RFIM-2} and other related models \cite{hysteresis-RA-1,hysteresis-RA-2}, as one tunes disorder variance continuously.    Another class of system is depinning of elastic manifolds inside a random potential \cite{review:Fisher,review:Kardar}.  The physics of the pinned phase, however, has not been systematically explored, largely due to the lack of appropriate formalism and language.   

Nevertheless, the general macroscopic theory of rate independent hysteresis is largely undeveloped.  In the author's opinion, such a theory should care less about microscopic mechanisms responsible for hysteresis and focus on the macroscopic, material independent aspects: How does a hysteretic system remember its history? What is the macroscopic meaning of ``memory''? How can memory be erased? Exploration of these issues not only is important to advance our understanding of nonequilibrium physics, it may also lead to new designs and applications in information science, material science and engineering.  
In this letter, we introduce such a macroscopic formalism involving semigroup theory, as well as a few exact results about rate independent hysteresis and return point memory.  Detailed discussions and proofs will be presented in a separated publication \cite{Xing-Semigroup-long}.


We shall only consider hysteresis systems driven by a scalar external field, for the sake of simplicity.  To develop a macroscopic approach to hysteresis, we first need a macroscopic description, or a ``name'', for each metastable state relevant to rate independent hysteresis.  For a system at thermal equilibrium, a  state is fully characterized by a few macroscopic parameter, such as volume, entropy, or their conjugate external field, such as pressure, temperature, etc.   For a disordered system at zero temperature, these parameters do not uniquely determines the current state.  This is of course the essence of hysteresis: the system remembers its past.   The first important observation is that for a given hysteresis dynamics, and a given initial state, a history of external field corresponds to a unique final state. {\em Therefore we can label a metastable state by the field history that we use to prepare this particular state}.   That is, the field history is the ``order parameter'' of hysteresis systems.  
Two important comments: 1) There are metastable states that are not reachable by any field history.  However, these states are not relevant to the study of rate independent hysteresis. 2) More importantly, there may be multiple field histories that correspond to the same state.  This is the case of hysteresis system with RPM.  We will show that the redundancy in description can be ``gauged out'' inside the framework of semigroup theory.   Study of this ``gauge symmetry'' help us to understand the basic algebraic structure of hysteresis phenomena with return point memory.  


\begin{figure}
\begin{center}
\includegraphics[width=3.5cm]{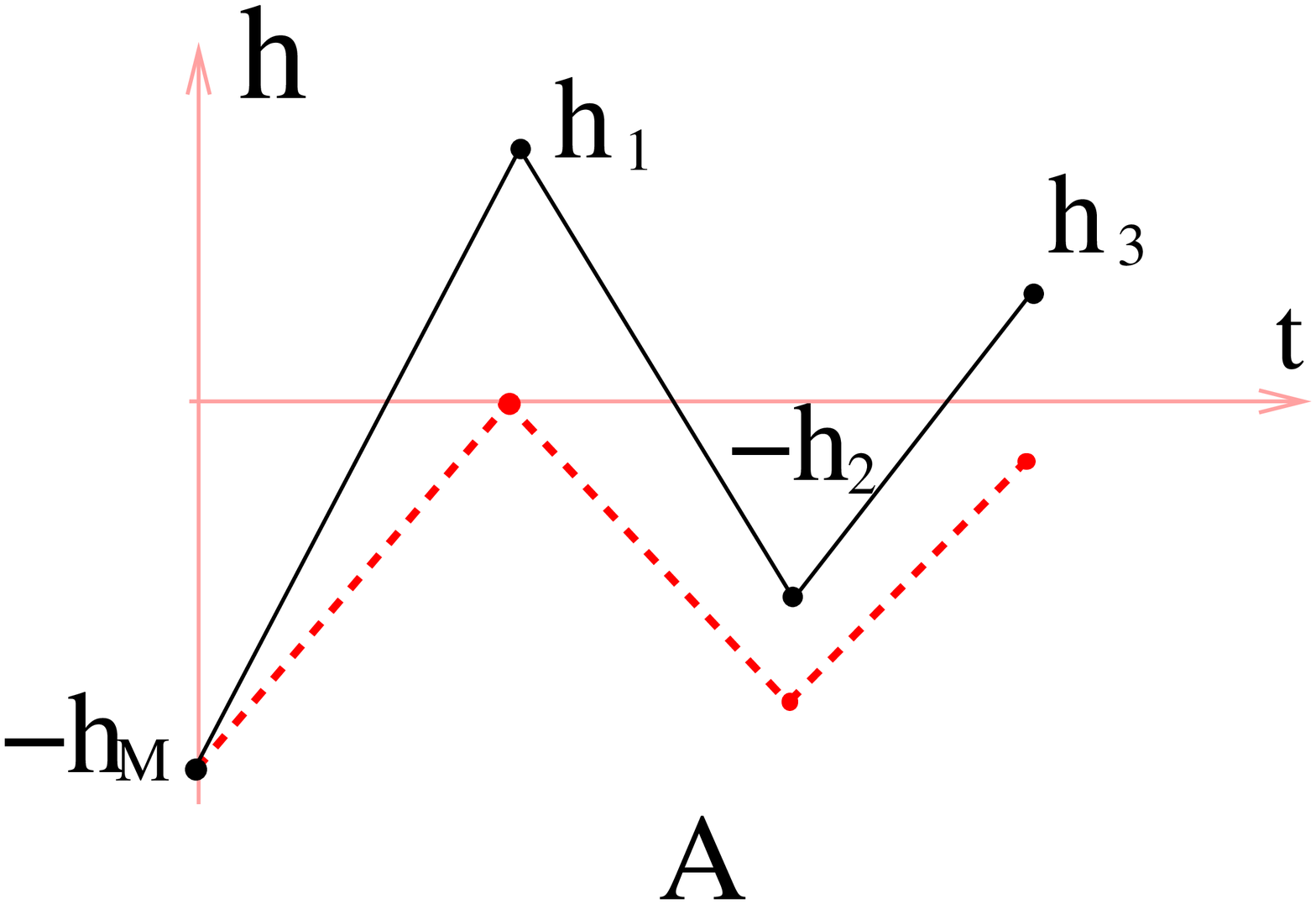}
\includegraphics[width=5cm]{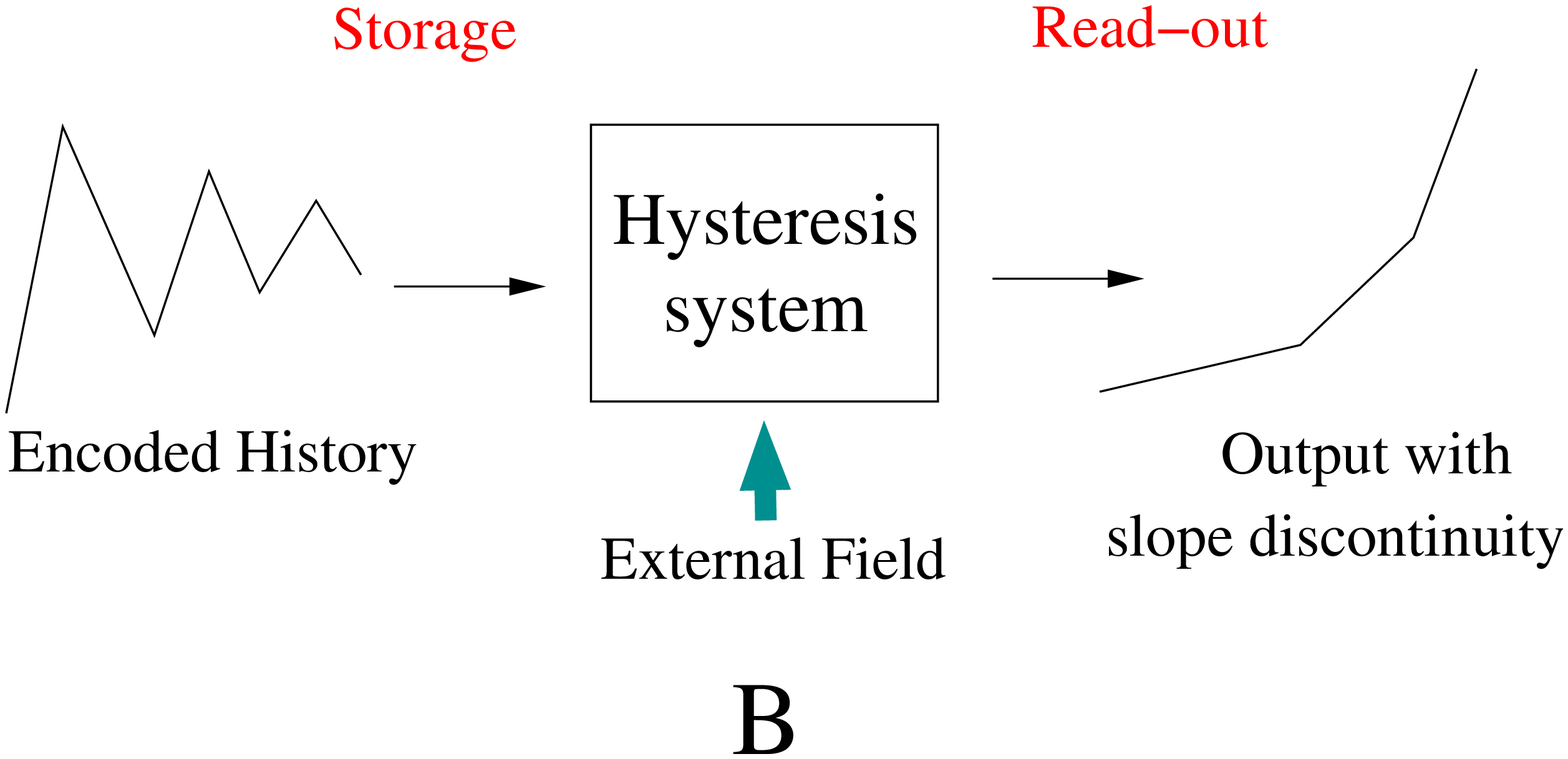}
\caption{{\small A: The external field (solid line) as a function of time determines a metastable state, for a given initial state.    Arbitrary time reparameterizations of this function represents the same field history.   The dashed line is another field history which is ``smaller'' than the first one, according to the relation $\geq$ defined in the history space.  \,\,
B: New mechanism for information storage and read-out using rate independent hysteresis. }}
\label{history-1}
\end{center}
\vspace{-8mm}
\end{figure}

Let us start from some initial metastable state at a field value $h_0$.   Very often we choose the initial state to be one of two saturated states $|+\rangle$ or $|-\rangle$, at the positive or negative saturation fields, $h_M$ or $-h_M$ respectively.  \footnote{In a spin model, all spin points up or down in these two saturation states.  However, the choice of these two saturation states is not essential for the semigroup theory.  The author thanks Jim Sethna for suggesting this point. }  We may introduce two type of operators,  $\eta(\Delta h)$, which increase the field by $\Delta h$, and $\xi(\Delta h)$, which decrease the field by $\Delta h$.   A metastable states reached by a particular hysteresis process then can be represented as a string of operators $\eta$ and $\xi$ acting on the initial state.   For example, for the history shown in Fig.~\ref{history-1}A, we start from the negatively saturated state $|-\rangle$ at field $-h_M$ and raise, lower, and raise the field, end up with a state with final field value $h_3$.  This state can be represented as 
$| \psi \rangle =  \eta(h_3 + h_2) \xi(h_2 + h_1) \eta(h_1 + h_M) 
| -\rangle $.  \footnote{We adopt the Dirac notation in quantum mechanics, where time propagates from right to left. }
The string of operators $\eta(h_3 + h_2) \eta(h_2 + h_1) \eta(h_1 + h_M)$ then represents the particular field history we used to prepare the metastable state $|\psi\rangle$.   Note that 1) raising or lowering the field by $\delta h$ always changes the metastable states, no matter how small $\delta h$ is; 2) The field history and the initial state determines at which field the final state is metastable.  For example, the state shown in left panel of Fig. \ref{history-1} is metastable at field 
$-h_M + (h_1+ h_M) - (h_2 + h_1) + (h_3 + h_2) = h_3.$
We can naturally put two histories together and form a product.  The product of $\eta(\Delta h_1)$ and $\xi(\Delta h_2)$ is just $\xi(\Delta h_2) \eta(\Delta h_1)$, which amounts to first increase field by $\Delta h_1$, then decrease it by $\Delta h_2$.  This is not the same as $\eta(\Delta h_1 - \Delta h_2)$, as the dynamics is irreversible.  It is also different from $\eta(\Delta h_1) \xi (\Delta h_2)$, exactly for the same reason.   Therefore the history product is not commutative.  On the other hand, the product of $\eta(h_1)$ and $\eta(h_2)$ is just $\eta(h_1 + h_2)$, which follows from adiabaticity.  Product of more complicated field histories can be recursively defined.   This product is clearly associative, i.e. $A(BC) = (AB)C$.  Therefore the set of all field histories forms a semigroup \cite{semigroup:Howie,Semigroup:Ljapin}, which we shall call the {\em general history semigroup}.   Now take a state $A |\psi_0 \rangle$ obtained by applying a field history $A$ onto the initial state $|\psi_0 \rangle$. Suppose we want to do hysteresis experiment by applying operation $B$ onto the system, the new state of the system after this experiment is simply given by $B A |\psi_0 \rangle$.   That is, {\em the rate independent hysteresis dynamics is simply represented as the product (concatenation) of histories}.   This strongly suggests that semigroup theory is {\em the natural language} for rate independent hysteresis.  

Almost all of the hysteresis models studied up to now possess a very important property, that is, return point memory (RPM,  or {\em wipe out property}): If one raises the external field by $h$, then lowers it by $h$, and finally raises it again by $h$, then end result on the system is identical to that if one raises the field by $h$ and does nothing else.  RPM implies the closeness of hysteresis subloops regardless of the initial state. 
Sethna and collaborators \cite{Sethna-RPM} discovered a deep result about RPM: If there is a partial order in the configuration space and the hysteresis dynamics respects this order, i.e. no passing, then the system must exhibit return point memory.  Qualitatively speaking, a partial order is a binary relation $\geq$ in the configuration space that allows one to compare certain pairs of state without inconsistency.  It has to satisfy: 1) {\em reflexivity}: $a \geq a$; 2) {\em antisymmetry}:  if $a \geq b$ and $b \geq a$, then $a = b$; 3) {\em Transtivity}: if $a \geq b$ and $b \geq c$ then $a \geq c$.   
A dynamics in the configuration space satisfies the no passing rule, if it respects the partial order.  That is, if we have two initial states $|\psi_1(0) \rangle \geq |\psi_2(0) \rangle$, which are driven by two field histories $h_1(t) \geq h_2(t)$ respectively.  Then for arbitrary time $t>0$, we have $| \psi_1(t)\rangle \geq |\psi_2(t)\rangle$.  
Qualitatively, RPM means that a system starting behind and driven by a lower field can not pass another system starting in front and driven by a higher field.   This no-passing property was first explored by Middleton \cite{Alan:nopass}, in the setting of driven charge density waves.  It was subsequently realized that it applies to a wide class of zero temperature driven systems. 


The relation between partial order and return point memory discovered by Sethna {\it et al} is elegant, deep, and general.  It illustrates how general properties of hysteresis follow from a few essential features of the dynamics.  One naturally wonders whether the converse is true: If RPM holds, does it imply a partial order in the configurational space that is respected by the dynamics?   As a special case, recently Narayan {\it et al} \cite{Narayan-RPM} found that 1D antiferromagnetic Ising chain shows RPM if starts from saturation state, but not if starts from a generic state.    Even though the dynamics does not respect the trivial partial order defined by spin orientation, Narayan found a hidden partial order that satisfies no passing rule.  A general answer to this question can be obtained from the semigroup approach introduced in this work.  

Return point memory is represented by the following identities in the general history semigroup: 
\begin{eqnarray}
\xi(h)\eta(h) \xi(h) =  \xi(h), 
\hspace{4mm}\eta(h)\xi(h)\eta(h) = \eta(h).
\label{RPM}
\end{eqnarray}
It defines an equivalence relation in the general history semigroup.  All histories equivalent to each other under Eq.~(\ref{RPM}) correspond to the same metastable state.   The history semigroup for hysteresis system with RPM is then the quotient semigroup over the equivalence Eq.~(\ref{RPM}).   The description of metastable state in terms of field history therefore contains redundancy: {\em return point memory is  a gauge symmetry in the history space}.

We can define a binary relation $\geq$ in the history space: let $A$ and $B$ be two field histories, which are strings of operators $\eta$'s  and $\xi$'s.   We define $A \geq B$ (`` $A$ is larger than $B$'') if, 
when starting from the same initial field, there exists a time parameterization of these field histories such that $A(t) \geq B(t)$ for all $t$.  An example is shown in Fig.~\ref{history-1}A.  The relation $\geq$ is clearly reflective and transitive.  More importantly, one can prove that if the return point memory Eq.~(\ref{RPM}) holds, then the relation $\geq$ is also antisymmetric \cite{Xing-Semigroup-long}, i.e. it is a partial order.  Furthermore, one trivially see this partial order is respected by the hysteresis dynamics, i.e. no passing rule holds: If $A\geq B$, and $C \geq D$, then $AC \geq BD$ by definition of the relation $\geq$.     Combining this result with that of reference \cite{Sethna-RPM}, we arrive at the following basic theorem of rate independent hysteresis: 

{\bf Theorem (RPM and partial order)}:  a rate independent hysteresis system exhibits return point memory if and only if there exists a partial order in the space of all reachable metastable states that is respected by the dynamics (i.e. no-passing).   

We emphasize that the partial order $\geq$ is defined only in the history space, which is in one-to-one correspondence to the set of {\em metastable states that are reachable by rate independent hysteresis processes}.   This is a subspace of the whole configurational space.  \footnote{One recent work by Dahmen {\it et al} \cite{Dahmen-RPM-GS} shows that the ground state of a random field Ising model is no-passing with respect to the change of external field.  However, since it is not clear whether the ground state is reachable by rate independent hysteresis process, our semigroup approach can not say anything about its properties. } It may not be related to the naive partial order (defined by spin orientations in spin models for example) in any apparent fashion.  Neither we should not expect that this partial order can be extended to the whole configurational space.  One recent example is given by Narayan {\it et al} \cite{Narayan-RPM}.

The one-to-one correspondence between field histories and metastable
states reachable by a hysteresis process may provide a novel mechanism of information storage.   We can imagine to encode some data into the extremal values of a field history in a rate independent fashion, which is in turn applied to a zero temperature hysteretic system, such as a disordered magnet.  This process imprints the data sequence onto the hysteresis system.  One important question is then how to extract informations stored in the system.   For a system with return point memory, there is a simple mechanism achieving this goal:    
One applies a field which is monotonically increasing over time.    Whenever the field passes an extreme value of the original field history, one should observe a slope discontinuity in the input-output curve.   This mechanism, schematically illustrated in Fig. \ref{history-1}B, allows one to recover all the local maximums of the original field history.   Further research along this direction may lead to feasible applications of hysteresis properties in information science and technology.   

There are other problems that we need to worry about.  The information stored in a hysteresis system may get erased accidently by noise.   On the other hand, if we want to imprint new information onto a system, we want to make sure that the old information can be completely erased.   Again, these questions can be properly addressed using semigroup theory.  We define another binary relation $\rhd$ ({\em erasing}) in the history space:  A history $A$ erases another history $B$, or $A \rhd B$, if and only if $AB = A$.  
We see that the field history $B$ has no effect if it is followed by the history $A$, that is, $B$ is erased by $A$.   The following two results follow from basic properties of semigroup: 
\begin{itemize} 
\item All histories that erase the history $A$ form a subsemigroup, which we call the eraser subsemigroup of $A$. 
\item All histories that are erased by the history $A$ form a subsemigroup, which we shall call the erased subsemigroup of $A$.  
\end{itemize}
The structures of eraser semigroup and erased semigroup of a given history should be problems of fundamental importance in the general theory of hysteresis, since they charcterize the macroscopic meaning of memory.   For a general hysteresis system without RPM, one can show that the only way to erase a history is to raise or lower the field to saturation.  On the other hand, sweeping the field to saturation clearly erase all field histories.  For system with RPM, the author managed to completely characterize the structures of the eraser semigroup and the erased semigroup of a special subclass of field history.  Due to limit of space, results are not presented here but will be published elsewhere \cite{Xing-Semigroup-long}.  


\begin{figure}
\begin{center}
\includegraphics[width=4cm]{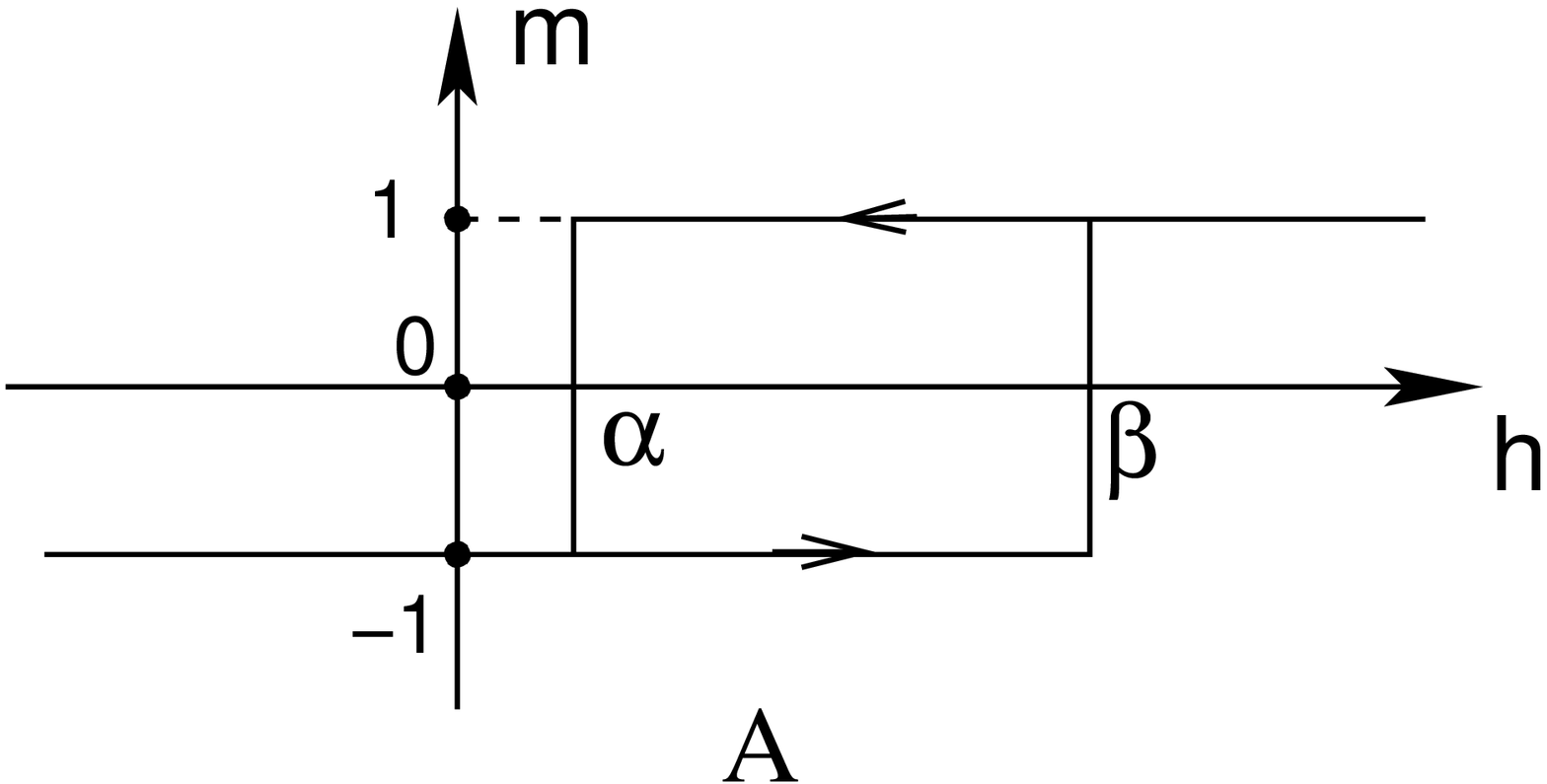}
\includegraphics[width=4cm]{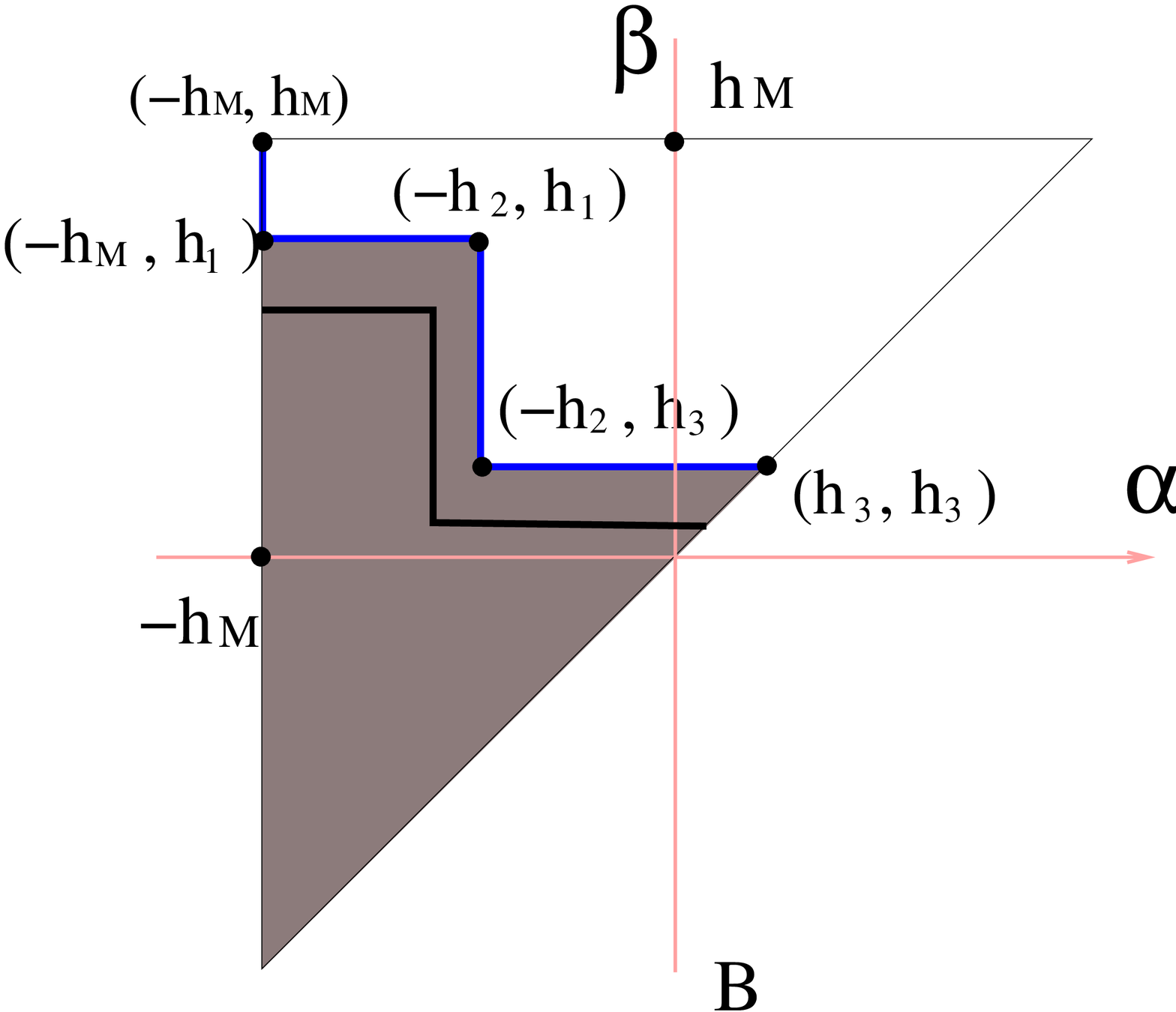}
\caption{{\small   A: Input-output relation for a hysteron. $h$ is the external field, $m$ is the ``magnetization''. 
B: Staircase representation of Preisach model and any hysteresis system with RPM. The partial order of field histories is also represented graphically as the relation ``above'' and ``below'' between staircases. }}
\label{staircase}
\end{center}
\vspace{-8mm}
\end{figure}

The Preisach model \cite{ref:Preisach,book:Mayergoyz} is one of the most successful phenomenological models for rate independent hysteresis.  It models hysteresis system as a superposition of many independent hysterons, i.e. basic units of hysteresis  with two switch fields $\alpha \leq \beta$, as illustrated in Fig.~\ref{staircase}A.   A hysteron exhibits bi-stability when $\alpha \leq h \leq \beta$. It flips up (its value changes from -1 to 1) when the external field passes $\beta$ from below, and flips down (its value changes from 1 to -1) when the field passes $\alpha$ from above.   The Preisach model admits a partial order (defined by the orientation of all hysterons) and no-passing rule, from which the return point memory follows.   Mayergoyz \cite{RPM:Mayergoyz} proved that if a hysteresis system satisfies return point memory and congruence property, then it can be represented by the Preisach model. 

%
The Preisach model admits an extremely useful graphic representation as shown in Fig.~\ref{history-1}B \cite{comment-staircase}: In the plane of two switch fields $\alpha$ and $\beta$, every point inside the triangle represents a hysteron.    A metastable state reachable by hysteresis processes starting from saturation states is then represented by a staircase running from one of the catheti to the hypotenuse.  All hysteron above/below the staircase are turned down/up.   For a given distribution of hysterons, this graphic representation constitutes a convenient starting point for the calculation of hysteresis loops \cite{book:Mayergoyz} in the input-output plane.  This staircase representation also gives an intuitive meaning to the partial order: a metastable state $A$ is larger than another one $B$ if and only if the staircase of $A$ is one the top of $B$ everywhere.    Finally the hysteresis dynamics of raising/lowering the field is represented by the deformation of the staircase to the above/left \cite{comment-staircase}.  

We make another observation that the staircase is a faithful representation of field history for any hysteresis system with RPM.  Using this graphic representation, one can show the following {\bf variational principle of rate independent hysteresis with RPM}:   Let $|\psi\rangle$ be an arbitrary initial state that is metastable at field value $h_0$.  Then the state $\eta(h) |\psi\rangle$ is the lower bound of all states that are 
1) metastable at field value $h_0 + h$, and 2) larger than $|\psi\rangle$.  
Similarly the state $\xi(h) |\psi\rangle$ is the upper bound of all states that are 1) metastable at field value $h_0 - h$, and 2) smaller than $|\psi\rangle$.  
This variational principle completely defines the hysteresis dynamics, just like the Hamiltonian variational principle completely defines the classical Lagrange dynamics.  It is amazing that an irreversible process such as rate independent hysteresis can be formulated from a variational principle.  

What is the difference between the Preisach model and the semigroup theory of hysteresis with RPM?  The semigroup theory addresses the most general properties of rate independent hysteresis that are {\em independent of microscopic details}.   The semigroup theory itself does not give prediction of the shapes of hysteresis loops.  For the latter purpose we need some {\it constitutive relations}.   In the Preisach model, these are given by the independent-hysteron assumption and the hysteron distribution.   Naturally different materials are characterized by different constitutive relations.  In a more general version of Preisach model, it is perceivable to give up the independent hysteron hypothesis and consider (weak) interactions.   This has indeed been studied by Mayergoyz and others \cite{book:Mayergoyz,book:Bertotti}.   It should be emphasized that disentanglement of the material-independent hysteresis properties from the materials-dependent ones is an important step towards a better understanding of hysteresis theory, and this can be  achieved by the semigroup approach.  Also further study of the variational principle may lead to discovery of more systematic approximation methods for the calculation of material properties.  It may also help us identify the connection between lattice models such as RFIM and more phenomenological models such as Preisach models.   These will be explored in the future.  



\begin{acknowledgments}
The author acknowledges financial support from the American Chemical Society under grant PRF 44689-G7, as well as stimulating discussions with Alan Middleton, Jim Sethna, and Chris Moore.  
\end{acknowledgments}



\end{document}